\documentclass[doublecol]{epl2}

\begin{document}

\title{Domain wall structure in magnetic bilayers with perpendicular anisotropy}
\shorttitle{Domain wall structure in magnetic bilayers with perpendicular anisotropy} 

\author{Amandine Bellec\inst{1}  \and Stanislas Rohart\inst{1}\thanks{E-mail: \email{rohart@lps.u-psud.fr}} \and Michel Labrune\inst{2} Jacques Miltat\inst{1} \and Andr\'e Thiaville\inst{1}}
\shortauthor{A. Bellec \etal}

\institute{
  \inst{1} Laboratoire de Physique des Solides, CNRS, Universit\'e Paris Sud, UMR 8502, 91405 Orsay Cedex, France\\
  \inst{2} Laboratoire PMTM, Institut Galil\'ee, CNRS, Universit\'e Paris-13, UPR 9001, 93430 Villetaneuse, France
}
\pacs{75.60.Ch}{Domain walls and domain structure}
\pacs{75.70.-i}{Magnetic properties of thin films, surfaces, and interfaces}
\pacs{68.37.Ef}{Scanning tunneling microscopy (including chemistry induced with STM)}

\abstract{We study the magnetic domain wall structure in magnetic bilayers (two ultrathin ferromagnetic layers separated by a non magnetic spacer) with perpendicular magnetization. Combining magnetic force and ballistic electron emission microscopies, we are able to reveal the details of the magnetic structure of the wall with a high spatial accuracy. In these layers, we show that the classical Bloch wall observed in single layers transforms into superposed N\'eel walls due to the magnetic coupling between the ferromagnetic layers. Quantitative agreement with micromagnetic calculations is achieved.}

\maketitle

\section{Introduction}

Magnetic domain walls (DW) are attracting a sustained interest both from a fundamental perspective and for applications, especially in the context of innovative magnetic memories~\cite{allwood2005,parkin2008}. The aim is to control the DW motion using magnetic fields or spin polarized currents. The dynamics of these walls is intrinsically linked to their micromagnetic structures, which impact their mobility and the so-called Walker velocity (the critical field or current above which the DW magnetization precesses when moving)~\cite{schryer1974,mougin2007}. In magnetic films with planar magnetization, the DW are generally large and are well characterized using classical tools such as magnetic force microscopy (MFM)~\cite{garcia2001}, scanning electron microscopy with spin polarization analysis (SEMPA)~\cite{berger1992} or x-ray photoemission electron microscopy (X-PEEM)~\cite{laufenberg2006}. On the contrary, in materials presenting a perpendicular magnetic anisotropy (PMA), the high magnetic anisotropy leads to narrow DW (width of few tens of nanometers or even less), which are difficult to measure and only the domains are observed~\cite{pommier1990,allenspach1990,bochi1995}. Based on micromagnetic arguments, in ultrathin films (thickness below the exchange length) the DW are generally thought to be Bloch walls~\cite{bloch1932,neel1955,bochi1995,sobolev1998}, which minimize the wall internal dipolar energy. However, this fact has never been explicitly proved by direct observation. This lack of evidence is surprising as these materials are widely studied, since they present original phenomena such as creep in small field~\cite{metaxas2007} and since their DW are thought to be easily moved using spin polarized current.

In this Letter, we present a study of DW in magnetic bilayers (two ferromagnetic layers separated by a non magnetic spacer) with perpendicular magnetization. Such structures have been recently studied and the dynamics under field or current has been shown to be original due to the coupling between the two layers~\cite{metaxas2010}. In order to gain precise information on their structure, we use a combination of MFM and ballistic electron emission microscopy (BEEM), a recent technique that allows high spatial resolution magnetization imaging. With the help of complementary micromagnetic calculations, we show that, contrary to intuition, the DW turn out to be N\'eel walls, due to the magneto-static coupling between the two ferromagnetic layers. Beyond this result, this study is also the first demonstration that BEEM can be applied to magnetic samples with perpendicular magnetization.

\section{Experimental details}

For the study of domain wall structure, high resolution magnetic imaging techniques have to be used. We have chosen to use the ballistic electron emission microscope (BEEM)~\cite{rippard2000}, a STM based technique with a theoretical resolution of few nanometers~\cite{prietsch1995}. One advantage of this technique is that it is not restricted to samples grown and observed in a ultra-high vacuum (UHV) environment, a fact that, e.g. enables complementary magnetic force microscopy (MFM) measurements. However, the BEEM technique also imposes some restrictions on the sample. In order to get enough signal, the magnetic layers have to be rather transparent to hot electrons in the 1-2~eV energy range. This forbids using Pt/Co multilayers, which is the archetypical example of multilayers with PMA, as the hot electron attenuation in Pt is very strong (more than 10 times higher than in Au~\cite{zhukov2006}). Instead, we use Au/Co multilayers, which provide similar magnetic anisotropies as Pt/Co multilayers but with a much better hot electron transparency.

The samples have been grown using electron-beam evaporation in UHV (base pressure $2\times10^{-10}$~mbar) on a hydrogenated $n$ type Si(111) substrate. The substrate has been cleaned and prepared using wet chemistry following the procedure described in Ref.~\cite{kaidatzis2008}, and displays atomically flat terraces prior to metal deposition. The first 5~nm thick Au layer ensures the formation of a high quality Schottky diode (barrier height $\Phi_B=0.8$~eV, ideality factor $<1.1$), which is a necessary condition for BEEM experiments. The following layers are Co(1.6~nm)/Au(5.0~nm)/Co(1.4~nm)/Au(5.0~nm) (the last layer is a capping layer which ensures the stability of the sample outside the UHV environment). The effective magnetic anisotropy is expected to be rather low (but positive) due to the rather large Co thickness. The advantage is that the domain wall energy is expected to be small enough to allow for the nucleation of small domains, thus facilitating BEEM experiments.

The BEEM experiments have been conducted using a modified Omicron-1 STM~\cite{thiaville2007} at room temperature and in a UHV chamber (different from the evaporation chamber). A magnetic field up to 60~Oe and perpendicular to the surface can be applied from outside the UHV chamber. For a positive tunneling voltage $V_T$, hot electrons are collected in the Si substrate and the transmission $T$ defined as the hot electron current $I_B$ to  tunneling current $I_T$ is measured. The MFM experiments have been conducted in ambient conditions on a Dimension~3000 Veeco AFM using home-made Co-Cr coated tips.

\section{Results}

\begin{figure}
  \includegraphics[width=0.9\columnwidth]{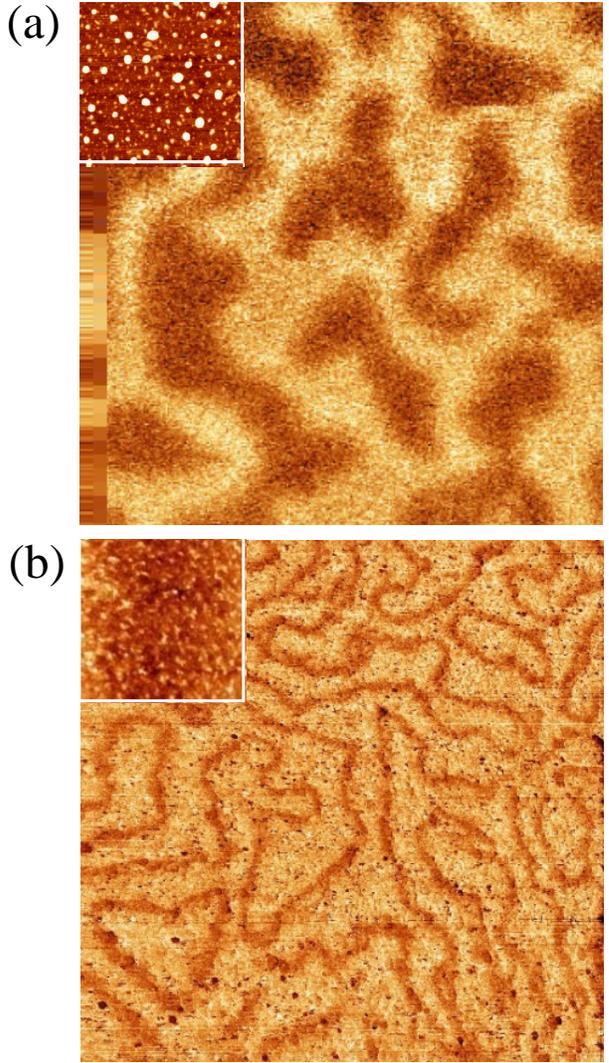}
  \caption{Images ($2 \times 2$~$\mu$m$^2$) of the  Co(1.6~nm)/Au(5~nm)/Co(1.4~nm) sample. (a)
  MFM phase image (inset topography)  (b) BEEM
  (Ballistic current) image (inset topography) (imaging conditions: $V_T = 1.5$~V,
  $I_T=20$~nA) }
  \label{fig:MFM_BEEM}
\end{figure}

The sample have first been studied using ``classical'' techniques, namely magneto-optical Kerr effect (MOKE) and MFM. The magnetic hysteresis loop measured in polar geometry shows that the sample indeed has a perpendicular magnetic anisotropy with a saturating field of 500~Oe, but does not have a 100~\% remanent magnetization ($M_R/M_S\approx20$~\%). This is an indication that in zero magnetic field, spontaneous nucleation of magnetic domains occurs. This phenomenon, observed in many samples like garnets, is attributed to a low DW energy, which allows minimizing the dipolar energy through the nucleation of small domains~\cite{kooy1960}. These domains have been revealed by MFM: the image (Fig.~\ref{fig:MFM_BEEM}a)
%
%
shows two types of domains with opposite magnetization (perpendicular to the surface), each domain brand occupying 50\% of the surface in zero magnetic field. The domain structure corresponds to segmented stripes with alternate magnetization direction but with a rather disordered structure. The average width of the stripes is $350 \pm 50$~nm. As only two contrast levels are seen in the MFM image, the magnetic state of the sample corresponds to either (i) one saturated layer and one layer with domains or (ii) a replication of the domains in the two layers. BEEM experiments have then been performed as the two hypotheses should give very different results, since BEEM probes the local alignment between the magnetization of the two layers. In the first case, we expect images similar to MFM images, whereas in the second case, we expect no contrast as the magnetization directions are always parallel to each other.

BEEM imaging has been first performed in zero magnetic field as shown in Fig.~\ref{fig:MFM_BEEM}b. The images are clearly different from MFM images but a clear contrast, uncorrelated with the STM topography is observed: dark lines (low local ballistic electron transmission) with a width of few tens of nanometers are present. One important characteristic is that the lines never cross each other and thus delimit closed areas, which is typical of a border. As explained before, such a result is unexpected. Assuming that the lines display the DW, these images correspond to a sample with a replication of the domains in the two layers: inside the domains the magnetization of the two layers are always parallel giving rise to a high transmission (bright areas).

\begin{figure}
  \includegraphics[width=0.9\columnwidth]{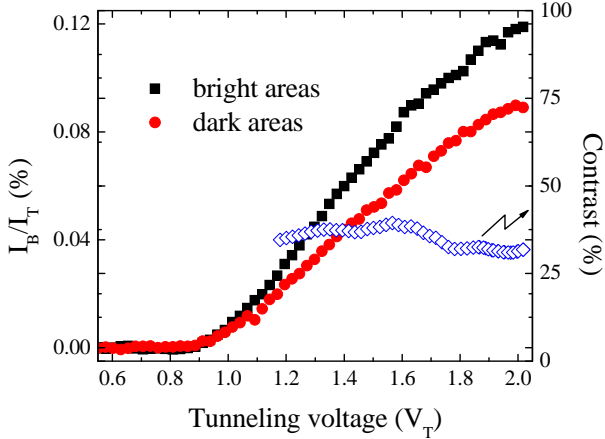}
  \caption{Hot electron transmission ($I_B/I_T$) versus tunneling voltage, recorded on the bright areas and on the dark areas. The transmission raises above 0.8~V, which corresponds to the Schottky barrier height at the Au/Si interface. The contrast evaluated with the help of equation \ref{eq:MC} is also presented. The tunneling current is $I_T=20$~nA.}
  \label{fig:spectro}
\end{figure}

To quantify the transmission and the magnetic contrast of our sample we performed BEEM spectroscopy measurements and compared the curves registered over the dark features and over the bright areas (Fig.~\ref{fig:spectro}).
%
Below 0.8~eV, no ballistic electrons are collected as their energy is lower than the Schottky energy barrier (this threshold is the same in both regions, which proves that the contrast does not arise from defects at the interface). Above 0.8~eV, we can clearly see the difference of transmission between the bright and the dark areas. We define the magnetic contrast $C$ (in percent) as
\begin{equation}
 C=100 \times \frac{T_{bright}-T_{dark}}{T_{dark}}\label{eq:MC}
\end{equation}
where $T_{bright}$ (resp. $T_{dark}$) is the transmission measured on a bright (resp. dark) area. This contrast is nearly energy independent, and is about $40 \pm 10$\%. Using the hot electron attenuation length obtained in previous work for Co~\cite{rippard2000,kaidatzis2008}, the magnetic contrast for our sample is estimated to range between 40~\% and 60~\% at 1.5V. This good agreement proves that the magnetization of the two layers are indeed locally anti-parallel in the dark zones.

\begin{figure}
  \includegraphics[width=0.9\columnwidth]{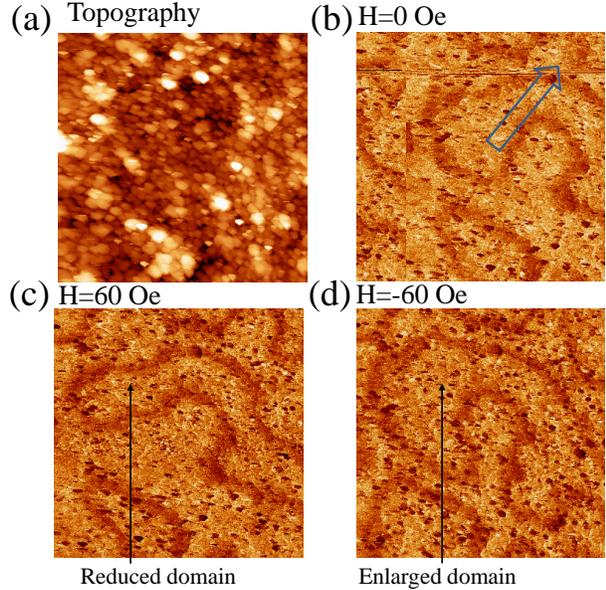}
  \caption{Evolution of the BEEM images with magnetic field (H).
  (a) STM topography, BEEM images at (b) H$=0$~Oe, (c) H$=60$~Oe and
  (d) H$=-60$~Oe ($V_T = 1.5$~V, $I_T=20$~nA, $1 \times 1$~$\mu$m$^2$). The arrow in (b) indicates the profile shown in Fig.~\ref{Fig:BEEMprof}.}
\label{fig:BEEMvsField}
\end{figure}

In order to prove that these dark lines are DW, we have also performed BEEM imaging under magnetic field, applied perpendicularly to the sample surface. Fig.~\ref{fig:BEEMvsField} compares ballistic current images recorded at the same position without magnetic field (b) and with a field of 60~Oe (c) and $-60$~Oe (d). Compared to H$=0$~Oe (Fig.~\ref{fig:BEEMvsField}b), the central domain is reduced for a positive magnetic field (Fig.~\ref{fig:BEEMvsField}c) while it is enlarged for a negative field (Fig.~\ref{fig:BEEMvsField}d). Thus, it has been possible to move the dark lines without changing their width, which is typical of DW. Comparing BEEM images with the sample topography (Fig~\ref{fig:BEEMvsField}a), it is possible to move one step further. Some defects (higher grains) can be observed in the topography and seem to act as pinning centers: the domain walls are ``linking'' some of these defects and only move in the defect free  zones. This incidentally proves that the sample microstructure determines the coercivity.
%

\section{Discussion: origin of the contrast}

\begin{figure}
  \includegraphics[width=0.9\columnwidth]{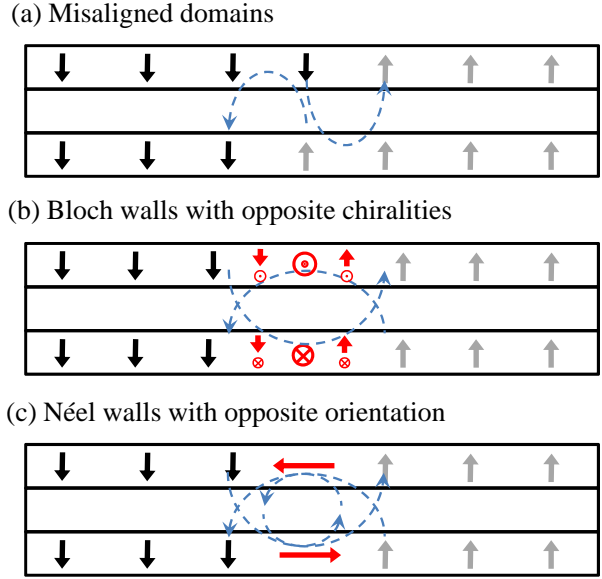}
  \caption{Possible configurations explaining the contrasts observed. (a) Misalignment of magnetic domains, (b) Bloch walls or
  (c) N\'eel walls with opposite chirality. The dashed lines sketch the flow of the stray field.}
  \label{Fig:Hypo}
\end{figure}

The origin of the anti-parallel alignment in the domain walls should now be explained. The interpretation is based on the dipolar coupling only, as the Au spacer thickness (5~nm) is sufficient to neglect any direct coupling~\cite{grolier1993}.  We can rule out that domains are misaligned (Fig.~\ref{Fig:Hypo}a), which could give rise to an AP alignment near the domains border. Indeed, the width of the observed lines is very regular and field independent (in the small field range $<100$~Oe). Moreover, the stray field around the domain walls tends to stabilize the alignment of the domain walls as depicted in Fig.~\ref{Fig:Hypo}a. We thus rather claim that the dark lines correspond to domain walls having opposite chirality in the two layers, leading to the local AP alignment.
%
In single ultrathin magnetic layers, DW are Bloch walls (helicoidal rotation of the spins in the DW plane) so that no magnetic charges are present, which minimizes the DW internal dipolar energy. These walls may have two chiralities (left or right handed). However, as no specific stray field is associated with these walls, they cannot couple to each other (neither ferro- nor antiferromagnetically). In our multilayer system, the probability to get opposite chiralities is then $\approx50$~\%. Since the size of domains observed by BEEM is in agreement with MFM observations, we know that all the walls are characterized by an AP alignment of the magnetization of the two layers.

Furthermore, it is easy to see that Bloch walls are unstable in this multilayer sample. As shown in Fig.~\ref{Fig:Hypo}b, at the center of one wall the stray field of the domains in the other layer is aligned along the $x$ direction (perpendicular to the wall plane). Since, at that position, the magnetization of Bloch walls is along the $y$ direction, this field creates a torque, which destabilizes the magnetization toward the $x$ direction. Due to this field, the walls may transform to N\'eel walls with a magnetization along the $x$ direction, as shown in Fig.~\ref{Fig:Hypo}c. More interestingly, the torques developed in each layer have an opposite direction, which leads to an opposite chirality in the two N\'eel walls. Moreover, as N\'eel walls are charged, an additional stray field, specific to the walls, is induced, which even strengthens the AP alignment of the two layer magnetization in the walls. This effect is at the origin of the contrast observed in the BEEM images. Such configurations can be reproduced easily using micromagnetic calculations for magnetic layers with a uniaxial perpendicular magnetic anisotropy $K$, with any value of parameters (i.e. thickness, anisotropy, magnetization...), provided that the magnetic anisotropy is larger than or equal to the shape anisotropy ($\frac{1}{2}\mu_0M_S^2$, with $M_S$ the layer magnetization) and that the thickness is below the exchange length.

\begin{figure}
  \includegraphics[width=0.9\columnwidth]{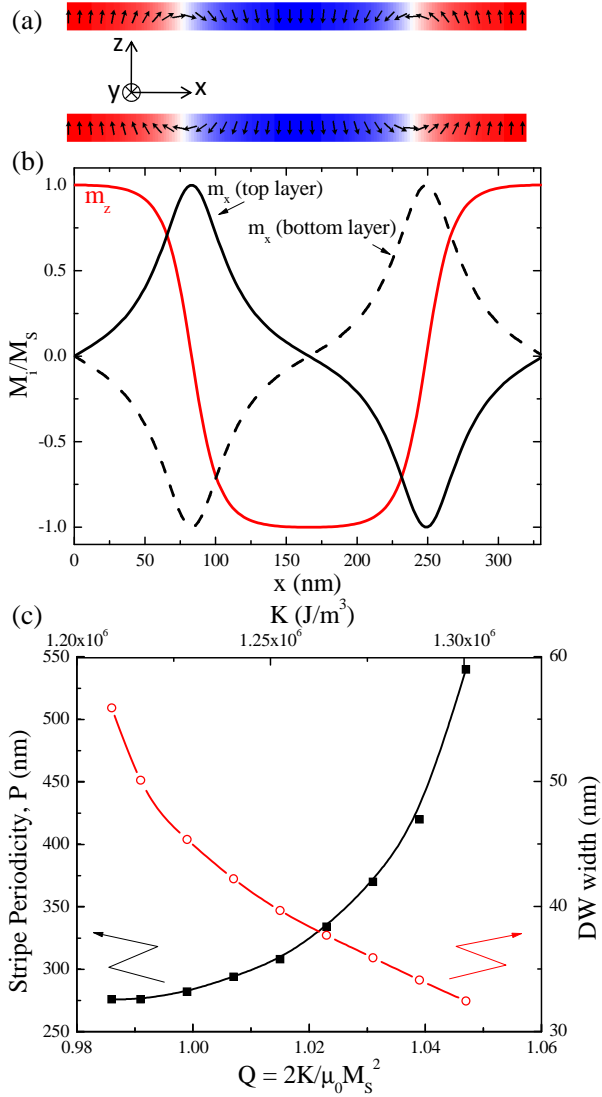}
  \caption{Micromagnetic calculation of a stripe network for two identical ferromagnetic layers (1.5~nm thick) separated by a non magnetic spacer (5~nm thick). (a) Schematic configuration of one stripe period (not to the scale) and (b) calculation result for $K = 1.26\times10^6$~J/m$^3$. (c) Stripe periodicity $P$ and  $DW$ width (see definition below) variation vs. quality factor $Q = 2K/\mu_0M_S^2$.}
  \label{Fig:simul}
\end{figure}

We now try to get a quantitative analysis of our images using numerical micromagnetic calculations. Using the bulk Co values for exchange interaction ($A=3\times10^{-11}$~J/m) and magnetization ($M_S = 1.4\times10^6$~A/m), the only unknown parameter is $K$ which is difficult to measure, since our sample is not in a saturated state in zero magnetic field. We have performed 2D calculations ($x,z$ plane), with infinite dimensions in the $y$ direction (along the DW axis) for a system with stripes. In our in-house code~\cite{labrune1994}, the sample periodicity in the $x$ direction is enforced. For a given $K$, the value of the period is then adjusted such as to minimize the energy density. The simulations reproduce well the model proposed in Fig.~\ref{Fig:Hypo}c with N\'eel walls with opposite chirality in the two layers (see Fig.~\ref{Fig:simul}a and b). In order to get a quantitative comparison, the goal is first to determine the value of $K$ that reproduces the experimental periodicity $P$ and compare the experimental and calculated DW profile. In a first approximation, we consider two identical layers (same thickness, same magnetic anisotropy). In the range where $K$ is of the same order of magnitude as the shape anisotropy, the period is a very good criterion, as $P$ depends strongly on $K$ as shown in Fig.~\ref{Fig:simul}c.
%

\begin{figure}
  \includegraphics[width=0.9\columnwidth]{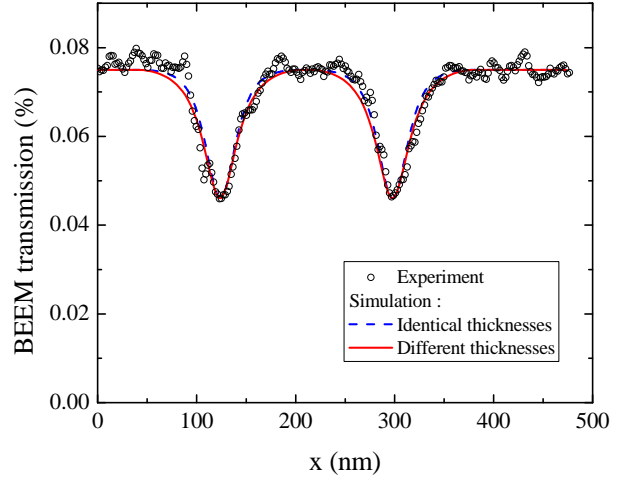}
  \caption{Comparison between the experimental profile (taken along the arrow in the BEEM image Fig~\ref{fig:BEEMvsField}b) and the calculations. The profile has been taken through two DW and thus shows two minima. For the calculated profiles, the distance between the two DW has been adjusted manually after the calculation so that to correspond exactly to this experimental case.}
  \label{Fig:BEEMprof}
\end{figure}

Experimentally, the periodicity (assumed to be at equilibrium) is $350 \pm 50$~nm so that, according to Fig.~\ref{Fig:simul}c the quality factor is permitted to range (see Fig.~\ref{Fig:simul}c) between 1.011 and 1.037 ($K =  1.24 - 1.28\times10^6$~J/m$^3$). From the calculated magnetization profile, the BEEM signal can now be simulated. Supposing that the two spin conduction channels in the spacer are independent, the BEEM signal is estimated to be~\cite{kaidatzis2008}
\begin{equation}
 T_{\mathrm{BEEM}}=\frac{T_P+T_{AP}}{2}+\frac{T_P-T_{AP}}{2}\stackrel{\rightarrow}{m_1}(x).\stackrel{\rightarrow}{m_2}(x),
\end{equation}
\noindent where $\stackrel{\rightarrow}{m_1}(x)$ and $\stackrel{\rightarrow}{m_2}(x)$ are the local magnetization orientations in the top and bottom layers. Across the DW, the BEEM signal shows a deep minimum corresponding to the anti-parallel alignment of the magnetization. The DW width, defined as the full width at half minimum of the BEEM signal, is shown in Fig.~\ref{Fig:simul}c. For $Q$ between 1.011 and 1.037, we estimate a DW width of $40\pm5$~nm, close to the experimental value ($43\pm5$~nm). In the experimental BEEM images, we have recorded a profile in order to compare with the calculation as shown in Fig.~\ref{Fig:BEEMprof} (note that several profiles on various images have been taken and they all show the same DW shapes).
The very good agreement between the experimental and calculated profile strongly supports the  magnetic structure model proposed here.
%

In the sample, the actual Co layers thicknesses are slightly different (respectively 1.4 and 1.6~nm). To better describe the sample reality, we now consider two different layers in the simulations. Since the magnetic anisotropy has an interfacial origin, the total magnetic anisotropy of a film with a thickness $t$ can be written as $K=K_V+2K_S/t$, where $K_V$ is the cobalt volume magnetocrystalline anisotropy (0.4~MJ/m$^3$~\cite{Beauvillain1994}) and $K_S$ is the surface anisotropy. Thus, the thicker film is expected to have a smaller anisotropy than the thinner one. Considering that the anisotropy found in the first calculations corresponds to the average of the anisotropies of the two films, we estimate $K_S = 0.66\pm0.02$~mJ/m$^2$ in good agreement with values found from the literature~\cite{Beauvillain1994}. The simulation using $K_S = 0.66$~mJ/m$^2$ and those different film thicknesses provides results almost identical to those presented above, but with a larger DW width in the thicker layer. The equilibrium period is found to be 340~nm in good agreement with the experiments, and the simulated BEEM profile reproduces very well the experimental one, as shown in Fig.~\ref{Fig:BEEMprof}.

\section{Conclusion}

In conclusion, we have shown that in a magnetic multilayer with perpendicular magnetization, the usual Bloch walls turn out to be unstable. A new type of wall is found consisting of one N\'eel wall in each layer, with opposite chirality. The details of these walls have been revealed with a high spatial resolution using the ballistic electron emission microscope. The simulations reproduce very well the observed features and show that this type of wall occurs in such layers for any magnitude of magnetic anisotropies, consistent with the existence of a perpendicular magnetization. The consequences of such DW structure modification on their dynamics ought to be detectable.

\acknowledgments
We gratefully acknowledge the help of T. Maroutian for e-beam evaporation and the loan of his deposition setup. Financial support was provided by C'Nano \^Ile-de-France (STCalib project).

\bibliographystyle{eplbib}
\bibliography{biblio}

\end{document}